\documentclass[11pt]{article}

\usepackage[english]{babel}
\usepackage{a4wide}
\usepackage{xspace}
\usepackage{graphicx,algorithm,algorithmic}
\usepackage{latexsym,amsmath,amssymb} 
\usepackage[usenames]{color,colortbl}
\usepackage{subfigure,epsfig}
\usepackage{rotating,lscape}
\usepackage{mdwlist}
\usepackage{setspace}
\usepackage{pstricks,pst-node,pst-plot}

\usepackage{amsfonts}
\usepackage{amsmath}
\usepackage{amsfonts}
\usepackage{amssymb}
\usepackage{url}
\newcommand{\MYTRUE}{\textsc{true}}
\newcommand{\MYFALSE}{\textsc{false}}

\newcommand{\kbw}{\ensuremath{k_\text{\tiny bw}}}

\definecolor{lightgray}{gray}{.85}

\begin{document}

\title{Iterative Beam Search for Simple Assembly Line Balancing with a Fixed Number of Work Stations}

\author{Christian~Blum \\
~\\
ALBCOM Research Group\\ 
Universitat Polit\`{e}cnica de Catalunya\\
c/ Jordi Girona 1-3, Campus Nord, Omega 112\\
08034 Barcelona (Spain)\\
{\sf cblum@lsi.upc.edu}}

\date{}

\maketitle

\begin{abstract}
The simple assembly line balancing problem (SALBP) concerns the assignment of tasks with pre-defined processing times to work stations that are arranged in a line. Hereby, precedence constraints between the tasks must be respected. The optimization goal of the SALBP-2 version of the problem concerns the minimization of the so-called cycle time, that is, the time in which the tasks of each work station must be completed. 

In this work we propose to tackle this problem with an iterative search method based on beam search. The proposed algorithm is able to obtain optimal, respectively best-known, solutions in 283 out of 302 test cases. Moreover, for 9 further test cases the algorithm is able to produce new best-known solutions. These numbers indicate that the proposed iterative beam search algorithm is currently a state-of-the-art method for the SALBP-2. 
\end{abstract}

\section{Introduction}

The class of problems known as assembly line balancing problems (ALBPs) concerns the optimization of processes related to the manufacturing of products via assembly lines. Their importance in the industrial world is shown by the fact that much research efforts have been dedicated to many different types of ALBPs during the past 50-60 years~\cite{Sal55:jie,GosGag89:alb}. The specific problem considered in this paper is the so-called simple assembly line balancing problem (SALBP)~\cite{BecSch2006:ejor}, a well-studied scientific test case. An assembly line is composed of a set of work stations arranged in a line, and by a transport system which moves the product to be manufactured along the line. The product is manufactured by executing a given set of tasks. Each of these tasks has a pre-defined processing time. In order to obtain a solution to a given SALBP instance, all tasks must be assigned to work stations subject to precedence constraints between the tasks. In the context of the SALBP, all work stations are considered to be of equal size. Moreover, the assembly line is assumed to move in constant speed. This implies a maximum of $C$ time units---the so-called \emph{cycle time}---for processing the tasks assigned to each work station. The SALBP has been tackled with several objective functions among which the following ones are the most studied ones in the literature:
\begin{itemize}
  \item Given a fixed cycle time $C$, the optimization goal consists in minimizing the number of necessary work stations. This version of the problem is refered to as SALBP-1. 
  \item Given a fixed number $m$ of work stations, the goal is to minimize the cycle time $C$. The literature knows this second problem version as SALBP-2.
\end{itemize}
The feasibility problem SALBP-F arises when both a cycle time $C$ and a number of work stations $m$ is given and the goal is to find a feasible solution respecting $C$ and $m$. In this work we will deal with the SALBP-2 version of the problem.

For what concerns the comparison between SALBP-1 and SALBP-2, much of the scientific work has been dedicated to the SALBP-1. However, also for the SALBP-2 exists a considerable body of research papers. An excellent survey was provided by~\cite{BecSch2006:ejor}. Approaches for the SALBP-2 can basically be classified as either \emph{iterative solution approaches} or \emph{direct solution approaches}. Iterative approaches tackle the problem by iteratively solving a series of SALBP-F problems that are obtained by fixing the cycle time. This process is started with a cycle time that is set to some calculated upper bound. This cycle time is then decremented during the iterative process, which stops as soon as no solution for the corresponding SALBP-F problem can be found. In contrast to these indirect approaches, direct approaches intend to solve a given SALBP-2 instance directly. 

Heuristic as well as complete approaches have been devised for the SALBP-2. Among the existing complete methods we find iterative approaches such as the ones proposed in~\cite{HacMagWee89:or,Sch99:book} but also direct approaches such as the ones described in~\cite{Sch94:orp,KleSch96:ejor}. Moreover, the performance of different integer programming formulation of the SALBP-2 has been evaluated in~\cite{PasFerGar07:iccsa}. The currently best-performing exact method is SALOME-2~\cite{KleSch96:ejor}. Surprisingly this exact method even outperforms the existing heuristic and metaheuristic approaches for the SALBP-2. The most successful metaheuristic approach to date is a tabu search method proposed in~\cite{SchVos96:ts}. Another tabu search proposal can be found in~\cite{Chi98:aor}. Other metaheuristic approaches include evolutionary algorithms~\cite{AndFer94:orsa,WatEtAl95:cep,Nea07:ijamt} and simulated annealing~\cite{Hen94:orp}. Finally, a two-phase heuristic based on linear programming can be found in~\cite{UguRacPap97:ejor}, whereas a heuristic based on Petri nets was proposed in~\cite{Kil10:ijamt}.

\paragraph{Contribution of this work.} Subsequently we propose to tackle the SALBP-2 by means of an iterative approach based on beam search, which is an incomplete variant of branch \& bound.\footnote{The interested reader may note that we proposed a very similar approach for a more general problem, the assembly line worker assignment and balancing problem (ALWABP)~\cite{BluMir11:alwabp}.} The resulting iterative beam search algorithm is inspired by one of the current state-of-the-art methods for the SALBP-1, namely Beam-ACO~\cite{Blu08:informs}. Beam-ACO is a hybrid approach that is obtained by combining the metaheuristic ant colony optimization with beam search. In this work we propose to use the beam search component of Beam-ACO in an iterative way for obtaining good SALBP-2 solutions. Our computational results show indeed that the proposed algorithm is currently a state-of-the-art method for the SALBP-2. It is able to obtain optimal, respectively best-known, solutions in 283 out of 302 test cases. Moreover, in further 9 cases the algorithm is able to produce new best-known solutions. 

\paragraph{Organization of the paper.} In Section~\ref{sec:salbp-2} we present a technical description of the tackled problem. Furthermore, in Section~\ref{sec:algo} the proposed algorithm is described. Finally, in Section~\ref{sec:results} we present a detailed experimental evaluation and in Section~\ref{sec:conclusions} we conclude our work and offer an outlook to future work.

\section{The SALBP-2}
\label{sec:salbp-2}

The SALBP-2 can technically be described as follows. An instance $(T,G,m)$ consists of three components. $T = \{1,\ldots,n\}$ is a set of $n$ tasks. Each task $i \in T$ has a pre-defined processing time $t_i > 0$. Moreover, given is a precedence graph $G=(T,A)$, which is a directed, acyclic graph with $T$ as node set. Finally, $m$ is the pre-defined number of work stations which are ordered from 1 to $m$. An arc $l_{i,j} \in A$ indicates that $i \in T$ must be processed before $j \in T$. Given a task $j \in T$, $P_j \subset T$ denotes the set of tasks that must be processed before $j$. A feasible solution is obtained by assigning each task to exactly one work station such that the precedence constraints between the tasks are satisfied. The objective function consists in minimizing the so-called cycle time. The SALBP-2 can be expressed in the following way as an integer programming (IP) problem.
\begin{equation}
  \mbox{\bf min } z
\end{equation}
subject to:
\begin{eqnarray}
\sum\limits_{s = 1}^{m} x_{is}   & =    & 1 \;\;\; \forall i \in T \\
\sum\limits_{s = 1}^{m} s x_{is} & \leq & \sum\limits_{s = 1}^{m} s x_{js} \;\;\; \forall j \in T, i \in P_j \\
\sum\limits_{i \in T} t_{i} x_{is}                  & \leq & z \;\;\; s = 1,\ldots,m \\
x_{is}                                               & \in  & \{0,1\} \;\;\; \forall i \in T, s = 1,\ldots,m \\
z                                                     & >    & 0  
\end{eqnarray}

This IP model makes use of the following variables and constants: $x_{is}$ is a binary variable which is set to 1 if and only if task $i \in T$ is assigned to work station $1 \leq s \leq m$. The objective function (1) minimizes the cycle time $z > 0$.\footnote{Note that we refer to the variable cycle time of the IP model as $z$, while fixed cycle times are denoted by $C$.} The constraints (2) ensure that each task $i \in T$ is assigned to a single work station $1 \leq s \leq m$. Constraints (3) reflect the precedence relationships between the tasks. More specifically, if task $j \in T$ is assigned to a work station $1 \leq s \leq m$, all tasks $i \in P_j$ must be assigned to work stations $1 \leq s^{\prime} \leq m$ with $s^{\prime} \leq s$. The constraints (4) ensure that the sum of the processing times of the tasks assigned to a work station $1 \leq s \leq m$ do not exceed the cycle time $z$. Note that this model is not necessarily the most efficient IP model for solving the SALBP-2. An evaluation of several different models can be found in~\cite{PasFerGar07:iccsa}.

\subsection{Solution Representation}

The following solution representation is used for the description of the algorithm as given in Section~\ref{sec:algo}. A solution $\mathcal{S}$ is an ordered list $\mathcal{S} = \langle S_1,\ldots,S_m\rangle$ of $m$ sets of tasks, where $S_i$ denotes the set of tasks that are assigned to the $i$-th work station. Abusing notation we henceforth call $S_i$ a work station. Note that for a solution $\mathcal{S}$ to be valid the following conditions must be fulfilled:
\begin{enumerate}
  \item $\bigcup_{i=1}^m S_i = T = \{1,\ldots,n\}$ and $\bigcap_{i=1}^m S_i = \emptyset$, that is, each task is assigned to exactly one work station.
  \item For each task $j \in S_i$ it is must hold that $P_j \subseteq \bigcup_{k=1}^{i} S_k$. This ensures that the precedence constraints between the tasks are not violated.
\end{enumerate}

\subsection{Reverse Problem Instances} 

The reverse problem instance $(T,G^r,m)$ with respect to an original instance $(T,G,m)$ is obtained by inverting the direction of all arcs of $G$. It is well-known from the literature~\cite{BecSch2006:ejor} that tackling the reverse problem instance may lead an exact algorithm faster to an optimal solution, respectively, may provide a better heuristic solution when tackled with the same heuristic as the original problem instance. Moreover, a solution $\mathcal{S}^r$ to the reverse problem instance $(T,G^r,m)$ can easily be converted into a solution $\mathcal{S}$ to the original problem instance $(T,G,m)$ as follows:
\begin{equation}
  S_i := S^r_{m-i+1} \quad \mbox{for } i = 1,\ldots,m 
\end{equation}

\section{Iterative Beam Search}
\label{sec:algo}

As mentioned in the introduction, the basic component of our algorithm for the SALBP-2 consists of beam search (BS), which is an incomplete derivative of branch \& bound~\cite{OwMor1988:ijpr}. Initially BS has especially been used in the context of scheduling problems (see, for example,~\cite{SabBay99:ejor,GhiPot05:ejor,ValAlv05:cie}). To date only very few applications to other types of problems exist (see, for example,~\cite{LeeWoo04:aca,AkeHifHal09:cor,BluBleLop09:lcs}). In the following we briefly describe how one of the standard versions of BS works. The crucial aspect of BS is the parallel extension of partial solutions in several ways. At all times, the algorithm keeps a set $B$ of at most $\kbw$ partial solutions, where $B$ is the so-called \emph{beam}, and $\kbw$ is known as the \emph{beam width}. At each step, at most $k_{\mbox{\tiny ext}}$ feasible extensions of each partial solution in $B$ are selected on the basis of greedy information. In general, this selection is done deterministically. At the end of each step, the algorithm creates a new beam $B$ by choosing up to $\kbw$ partial solutions from the set of selected feasible extensions. For that purpose, BS algorithms determine---in the case of minimization---a lower bound value for each extension. Only the maximally $\kbw$ best extensions---with respect to these lower bound values---are included in $B$. Finally, if any complete solution was generated, the algorithm returns the best of those. Note that the underlying constructive heuristic that defines feasible extensions of partial solutions and the lower bound function for evaluating partial solutions are crucial for the working of BS.

In the following we first present a description of the implementation of the BS component, before we describe the algorithmic scheme in which this BS component is used.

\subsection{The Beam Search Component}

The BS component described in this section---see Algorithm~\ref{algo:bs} for the pseudo-code---is the main component of the proposed algorithm for the SALBP-2. The algorithm requires a problem instance $(T,G,m)$, a fixed cycle time $C$, a beam width $\kbw$, and a maximal number of extensions $k_{\mbox{\tiny ext}}$ as input. Given a fixed cycle time $C$ and $m$ (the number of work stations) BS tries to find at least one feasible solution. As mentioned before, the crucial aspect of BS is the extension of partial solutions in several possible ways. At each step the algorithm extends each partial solution from $B$ in a maximum number of ways. More specifically, given a partial solution with $l-1 < m$ work stations already filled, an extension is generated by assigning a set of so-far unassigned tasks to the next work station $S_l$ such that the given cycle time $C$ is not surpassed and the precedence constraints between the tasks are respected (see lines 11--12 of Algorithm~\ref{algo:bs}). The algorithm produces extensions in a (partially) probabilistic way rather than in the usual deterministic manner.\footnote{This is done for avoiding a \emph{choice-without-replacement} process for which all possible work station fillings must be generated beforehand.} Each generated extension (partial solution) is either stored in set $B_{\mbox{\tiny compl}}$ in case it is a complete solution, or in set $B_{\mbox{\tiny ext}}$ otherwise (see lines 13--19 of Algorithm~\ref{algo:bs}). However, a partial solution is only stored in set $B_{\mbox{\tiny ext}}$ if it uses at most $m-1$ work stations, and if its $l$-th work station is different to the $l$-th work station of all partial solutions that are already in $B$. Finally, BS creates a new beam $B$ by selecting up to $k_{\mbox{\tiny bw}}$ solutions from set $B_{\mbox{\tiny ext}}$ of further extensible partial solutions (see line 22 of Algorithm~\ref{algo:bs}). This is done in function {\sf SelectSolutions}($B_{\mbox{\tiny ext}}$,$k_{\mbox{\tiny bw}}$) on the basis of a lower bound function LB$(\cdot)$. In the following we describe in detail the extension of partial solutions and the working of function {\sf SelectSolutions}($B_{\mbox{\tiny ext}}$,$k_{\mbox{\tiny bw}}$). 

\begin{algorithm}[t]
\caption{\label{algo:bs} Beam search}
\begin{algorithmic}[1]
  \STATE {\bf input:} an instance $(T,G,m)$, a fixed cycle time $C$, a beam width $\kbw$, and $k_{\mbox{\tiny ext}}$
  \STATE $l := 0$
  \STATE Initialization of an empty solution $\mathcal{S}$
  \STATE $B := \{\mathcal{S}\}$
  \STATE $B_{\mbox{\tiny compl}} := \emptyset$
  \WHILE{$B \not= \emptyset$}
    \STATE $B_{\mbox{\tiny ext}} := \emptyset$
    \STATE $l := l+1$
    \FOR{{\bf all} $\mathcal{S} \in B$}
      \FOR{$i=1,\ldots,k_{\mbox{\tiny ext}}$}
        \STATE $\mathcal{S}^{\prime} := \mathcal{S}$ \COMMENT{copy partial solution $\mathcal{S}$ into $\mathcal{S}^{\prime}$}
        \STATE $S^{\prime}_l :=$ {\sf ExtendPartialSolution($\mathcal{S}^{\prime},l,C$)} \COMMENT{see Algorithm~\ref{algo:fill-station}}
        \IF{solution $\mathcal{S}^{\prime}$ is complete (that is, all tasks are assigned)}
          \STATE $B_{\mbox{\tiny compl}} := B_{\mbox{\tiny compl}} \cup \{\mathcal{S}^{\prime}\}$
        \ELSE
          \IF{$l < m$ and $\mathcal{S}^{\prime}_l$ is different to the $l$-th work station of all other $\mathcal{S} \in B_{\mbox{\tiny ext}}$}
            \STATE $B_{\mbox{\tiny ext}} := B_{\mbox{\tiny ext}} \cup \{\mathcal{S}^{\prime}\}$
          \ENDIF
        \ENDIF
      \ENDFOR
    \ENDFOR
    \STATE $B \leftarrow ${\sf SelectSolutions}($B_{\mbox{\tiny ext}}$,$k_{\mbox{\tiny bw}}$)
  \ENDWHILE
  \STATE {\bf output:} If $B_{\mbox{\tiny compl}} \not= \emptyset$ the output is $\MYTRUE$, otherwise $\MYFALSE$
\end{algorithmic}
\end{algorithm}

\paragraph{Extending Partial Solutions.} The generation of an extension of a partial solution $\mathcal{S}^{\prime}$ with $l-1$ work stations already filled works as follows. Unassigned tasks are iteratively assigned to work station $S^{\prime}_l$ until the sum of their processing times is such that no other task can be added to $S^{\prime}_l$ without exceeding the given cycle time $C$. This procedure is pseudo-coded in Algorithm~\ref{algo:fill-station}. At each step, $T^{\prime}$ denotes the set of so-far unassigned tasks that may be added to $S^{\prime}_l$ without violating any constraints. The definition of this set of \emph{available tasks} is given in line 3, respectively 8, of Algorithm~\ref{algo:fill-station}.

\begin{algorithm}[t]
\caption{\label{algo:fill-station} Function {\sf ExtendPartialSolution}($\mathcal{S}^{\prime},l,C$) of Algorithm~\ref{algo:bs}}
\begin{algorithmic}[1]
  \STATE {\bf input:} A partial solution $\mathcal{S}^{\prime}$, the index $l$ of the work station to be filled, and the cycle time $C$
  \STATE $S^{\prime}_l := \emptyset$
  \STATE $T^{\prime} := \{ i \in T \mid i \notin \bigcup_{j=1}^{l} S^{\prime}_j, P_i \subseteq \bigcup_{j=1}^{l} S^{\prime}_j, t_{i} + c_{\mbox{\tiny rem}} \leq C\}$
  \STATE $c_{\mbox{\tiny rem}} := C$
  \WHILE{$T^{\prime} \not= \emptyset$}
    \STATE $j := ${\sf ChooseTask($T^{\prime},c_{\mbox{\tiny rem}}$)}
    \STATE $c_{\mbox{\tiny rem}} := c_{\mbox{\tiny rem}} - t_{j}$
    \STATE $T^{\prime} := \{ i \in T \mid i \notin \bigcup_{j=1}^{l} S^{\prime}_j, P_i \subseteq \bigcup_{j=1}^{l} S^{\prime}_j, t_{i} + c_{\mbox{\tiny rem}} \leq C\}$
    \STATE $S^{\prime}_l := S^{\prime}_l \cup \{j\}$
  \ENDWHILE
  \STATE {\bf output:} Filled work station $S^{\prime}_l$
\end{algorithmic}
\end{algorithm}

It remains to describe the implementation of function {\sf ChooseTask($T^{\prime},c_{\mbox{\tiny rem}}$)} of Algorithm~\ref{algo:fill-station}. For that purpose let us first define the following subset of $T^{\prime}$:
\begin{equation}\label{eq:sat}
  T^{\mbox{\tiny sat}} := \{i \in T^{\prime} \mid t_{i} + c_{\mbox{\tiny rem}} = C\} 
\end{equation}
This definition is such that $T^{\mbox{\tiny sat}}$ contains all tasks that \emph{saturate}, in terms of processing time, the $l$-th work station $S_l$. The choice of a task from $T^{\prime}$ is made on the basis of greedy information, that is, on the basis of values $\eta_i > 0$ that are assigned to all tasks $i \in T^{\prime}$ by a greedy function. The first action for choosing a task from $T^{\prime}$ consists in flipping a coin for deciding if the choice is made deterministically, or probabilistically. In case of a deterministic choice, there are two possibilities. First, if $T^{\mbox{\tiny sat}} \not= \emptyset$, the best task from $T^{\mbox{\tiny sat}}$ is chosen, that is, the task with maximal greedy value among all tasks in $T^{\mbox{\tiny sat}}$. Otherwise, we choose the task with maximal greedy value from $T^{\prime}$. In case of a probabilistic decision, a task from $T^{\prime}$ is chosen on the basis of the following probability distribution:
\begin{equation}
  \mathbf{p}(i) := \frac{\eta_i}{\sum\limits_{j \in T^{\prime}} \eta_j} \enspace, \forall i \in T^{\prime}
\end{equation}
For completing the description of function {\sf ChooseTask($T^{\prime},c_{\mbox{\tiny rem}}$)}, we must describe the definition of the greedy values $\eta_i$, $\forall i \in T$. In a first step a term $\gamma_i$ is defined as follows:
\begin{equation}\label{eq:gamma}
  \gamma_i := \kappa_1 \cdot \left(\frac{t_{i}}{C}\right) + \kappa_2 \cdot \left(\frac{\left|\mbox{Suc}^{\mbox{\tiny all}}_{i}\right|}{\mbox{max}_{1 \leq j \leq n} \left|\mbox{Suc}^{\mbox{\tiny all}}_{j}\right|}\right) \enspace, \forall i \in T
\end{equation}
Hereby, $\mbox{Suc}^{\mbox{\tiny all}}_{i}$ denotes the set of all tasks that can be reached from $i$ in precedence graph $G$ via a directed path. This definition combines two greedy heuristics that are often used in the context of assembly line balancing problems. The first one concerns the task processing times and the second one concerns the size of $\mbox{Suc}^{\mbox{\tiny all}}_{i}$. The influence of both heuristics can be adjusted via the setting of weights $\kappa_1$ and $\kappa_2$. In order to be more flexible we decided to allow for both weights a value from $[-1,1]$. This means that we consider for each heuristic potentially also its negation. Given the $\gamma_i$-values, the greedy values $\eta_i$ are then derived as follows:
\begin{equation}\label{eqn:aco-heuristic-information}
  \eta_i := \frac{\gamma_i - \gamma_{\mbox{\tiny min}} + 1}{\gamma_{\mbox{\tiny max}}} \;\;\; \forall \; i \in T \enspace,
\end{equation}
where $\gamma_{\mbox{\tiny min}}$, respectively $\gamma_{\mbox{\tiny max}}$, denote the minimum, respectively maximum, values of all $\gamma_i$. Interestingly, for obtaining well-working greedy values, parameters $\kappa_1$ and $\kappa_2$ have to be chosen in a problem-instance-dependent way. Tuning experiments are present in Section~\ref{sec:tuning}.

\paragraph{The Lower Bound Function.} The new beam $B$ is---at each step---chosen from $B_{\mbox{\tiny ext}}$. This choice is implemented by function {\sf SelectSolutions}($B_{\mbox{\tiny ext}}$,$k_{\mbox{\tiny bw}}$) of Algorithm~\ref{algo:fill-station}. First, the solutions in $B_{\mbox{\tiny ext}}$ are ranked with respect to increasing lower bound values LB($\cdot$). Then, the min$\{k_{\mbox{\tiny bw}},|B_{\mbox{\tiny ext}}|\}$ highest ranked partial solutions from $B_{\mbox{\tiny ext}}$ are selected. Let us denote by $\overline{T} \subseteq T$ the set of tasks that have not yet been assigned to work stations in partial solution $\mathcal{S}^{\prime}$. Then:
\begin{equation}
  \mbox{LB}(\mathcal{S}^{\prime}) = \left\lceil \frac{\sum_{i \in \overline{T}} t_i}{C} \right\rceil
\end{equation}
Note that this lower bound is inspired by splitting-based bounds for the one-dimensional bin-backing problem.

\subsection{The Algorithmic Scheme}
\label{sec:algo-scheme}

\begin{algorithm}[!t]
\caption{\label{algo:scheme} Iterative beam search (\textsc{Ibs}) for the SALBP-2}
\begin{algorithmic}[1]
  \STATE {\bf input:} an instance $(T,G,m)$
  \STATE $C :=$ {\sf DetermineStartingCycleTime}()
  \STATE $\kbw := 5$, $k_{\mbox{\tiny ext}} := 2$
  \STATE $success := \MYFALSE$
  \WHILE{{\bf not} $success$}
    \STATE $success :=$ {\sf BeamSearch}($(T,G,m),C,\kbw,k_{\mbox{\tiny ext}}$) \COMMENT{original instance}
    \IF{{\bf not} $success$}
      \STATE $success :=$ {\sf BeamSearch}($(T,G^r,m),C,\kbw,k_{\mbox{\tiny ext}}$) \COMMENT{reverse instance}
      \STATE {\bf if} {\bf not} $success$ {\bf then} $C := C + 1$ {\bf end if}
    \ENDIF
  \ENDWHILE
  \STATE $C := C - 1$
  \STATE $stop := \MYFALSE$
  \WHILE{{\bf not} $stop$}
    \STATE $success := \MYFALSE$
    \WHILE{time limit not reached {\bf and not} $success$}
      \STATE {\bf if} within 5\% of time limit {\bf then} $\kbw := 10$, $k_{\mbox{\tiny ext}} := 5$ {\bf else} $\kbw := 150$, $k_{\mbox{\tiny ext}} := 20$ {\bf end if}
      \STATE $success :=$ {\sf BeamSearch}($(T,G,m),C,\kbw,k_{\mbox{\tiny ext}}$) \COMMENT{original instance}
      \IF{{\bf not} $success$}
        \STATE $success :=$ {\sf BeamSearch}($(T,G^r,m),C,\kbw,k_{\mbox{\tiny ext}}$) \COMMENT{reverse instance}
      \ENDIF
    \ENDWHILE
    \STATE {\bf if} $success$ {\bf then} $C := C - 1$ {\bf else} $stop := \MYTRUE$ {\bf end if}
  \ENDWHILE
  \STATE $C := C + 1$
  \STATE {\bf output:} cycle time $C$
\end{algorithmic}
\end{algorithm}

The BS component outlined in the previous section is used by an iterative algorithmic scheme that is presented in Algorithm~\ref{algo:scheme}. Henceforth this algorithmic scheme is labelled iterated beam search (IBS). The first step consists in determining a starting cycle time $C$, which is computed in funcion {\sf DetermineStartingCycleTime}() of Algorithm~\ref{algo:scheme} as
\begin{equation}
  C := \mbox{max}\left\{\mbox{max}_{i \in T} \{t_i\},\left\lceil \frac{\sum_{i \in T} t_i}{m} \right\rceil  \right\} \enspace.
\end{equation}
The algorithm works in two phases. In the first phase (see lines 3-11 of Algorithm~\ref{algo:scheme}) the algorithm tries to quickly find a first cycle time $C$ for which a valid solution can be found. For this purpose BS is applied with the setting $\kbw = 5$ and $k_{\mbox{\tiny ext}} = 2$. Note that this setting was chosen after tuning by hand. Moreover, note that the first phase only takes a fraction of a second of computation time. This holds for all instances considered in Section~\ref{sec:results}. The second phase of the algorithm iteratively tries to find a valid solution for the next smaller cycle time. In this phase, the algorithm disposes over a certain time limit for each considered cycle time. Remember that the working of BS is partially probabilistic. Therefore, BS can repeatedly be applied to the same instance with potentially different outcomes. The first five percent of the above-mentioned time limit are spent by BS applications that use the setting $\kbw := 10$ and $k_{\mbox{\tiny ext}} := 5$. This is done with the intention of not wasting too much computation time, if not necessary. However, if BS is not able to solve the given cycle time with this setting, the remaining 95\% of the available time are spent by BS applications using the setting $\kbw := 150$ and $k_{\mbox{\tiny ext}} := 20$. With this setting BS is considerably slower. However, the probability of finding feasible solutions is much higher than with the setting described before. The second phase of the algorithm ends when the time limit has passed without having found a feasible solution for the considered cycle time.

\section{Experimental Evaluation}
\label{sec:results}

\textsc{Ibs} was implemented in ANSI C++, and GCC 3.4.0 was used for compiling the software. Experimental results were obtained on a PC with an AMD64X2 4400 processor and 4 Gb of memory. In the following we first describe the set of benchmark instances that we used for the experimental evaluation. Subsequently we present the tuning process that we conducted and the results of the proposed algorithm.

\subsection{Benchmark Instances}

We used the usual set of 302 benchmark instances from the literature. They can be obtained---together with information about optimal and best-known solutions---from a website especially dedicated to all kind of assembly line balancing problems maintained by Armin Scholl, \url{http://www.assembly-line-balancing.de}. Each instance consists of a precedence graph $G$ and a given number $m$ of work stations.  The benchmark set is composed of two subsets of instances, henceforth called {\sf Dataset1} and {\sf Dataset2}. {\sf Dataset1} consists of 128 instances based on 9 different precedence graphs with a number of tasks between 29 to 111. {\sf Dataset2} is composed of 174 instances based on 8 different precedence graphs with a number of tasks varying from 53 to 297. 

\subsection{Algorithm Tuning}
\label{sec:tuning}

During preliminar experiments we realized that parameters $\kbw$ and $k_{\mbox{\tiny ext}}$ have a rather low impact on the final results of \textsc{Ibs}. In other words it is easy to find a reasonable setting for these parameters quite quickly. Their setting dynamically changes during a run of the algorithm as specified in Section~\ref{sec:algo-scheme}. On the contrary, parameters $\kappa_1$ and $\kappa_2$ (see Eq.~\ref{eq:gamma}) have a high impact on the algorithms' performance. Remember that $\kappa_1$ is the weight of the greedy function concerning the task processing times, while $\kappa_2$ is the weight of the greedy function concerning the number of tasks that have to be processed after the task under consideration. As mentioned before, for both parameters we allowed values from $[-1,1]$. Instead of trying to find a good parameter setting for each instance, we decided for a tuning process based on precedence graphs, that is, we wanted to choose a single setting of $\kappa_1$ and $\kappa_2$ for all instances concerning the same precedence graph. For that purpose we applied a specific version of \textsc{Ibs} for all combinations of $\kappa_1, \kappa_2 \in \{-1.0,-0.9,\ldots,0.0,\ldots,0.9,1.0\}$ to all 302 instances. This makes a total of 441 different settings for each instance. The specific version of \textsc{Ibs} that we applied for the tuning process differs from \textsc{Ibs} as outlined in Algorithm~\ref{algo:scheme} in that lines 14-24 were replaced by a single, deterministic, application of beam search with $\kbw = 150$ and $k_{\mbox{\tiny ext}}=20$. This was done for the purpose of saving computation time. Based on the obtained results we chose the settings presented in Table~\ref{tab:tuning} for the different precedence graphs. It is interesting to note that, apart from a few exceptions, the greedy heuristic based on task processing times does not seem necessary for obtaining good results. 

\begin{table}[!t]
\caption{\label{tab:tuning} Values of parameters $\kappa_1$ and $\kappa_2$ for the final experiments.}
\begin{center}
\begin{tabular}{l|ll||l|ll} \hline
{\bf Graph} & $\kappa_1$ & $\kappa_2$ & {\bf Graph} & $\kappa_1$ & $\kappa_2$ \\ \hline
Arcus1      & 0.0        & 1.0        & Lutz2       & -0.1       & 0.9 \\
Arcus2      & -0.5       & 0.9        & Lutz3       & 0.0        & 1.0 \\
Barthol2    & 0.0        & 1.0        & Mukherje    & 0.0        & 1.0 \\
Barthold    & 0.0        & 1.0        & Sawyer      & 0.0        & 1.0 \\
Buxey       & 0.0        & 1.0        & Scholl      & 0.0        & 1.0 \\
Gunther     & -0.4       & 0.8        & Tonge       & -0.1       & 0.2 \\
Hahn        & 0.0        & 1.0        & Warnecke    & 0.0        & 1.0 \\
Kilbridge   & 0.0        & 1.0        & Wee-Mag     & 0.0        & 1.0 \\
Lutz1       & -0.1       & 0.9        &             &            &     \\ \hline
\end{tabular}						       
\end{center}
\end{table}

\begin{figure}[p]
\centering
\subfigure[Buxey, $m=7$]{
\label{fig:tuning:a}
  \includegraphics[width=6cm]{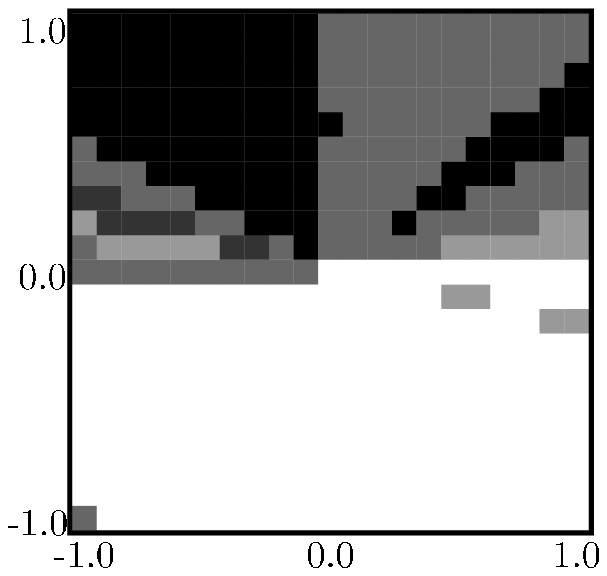}
}
\subfigure[Tonge, $m=16$]{
\label{fig:tuning:b}
  \includegraphics[width=6cm]{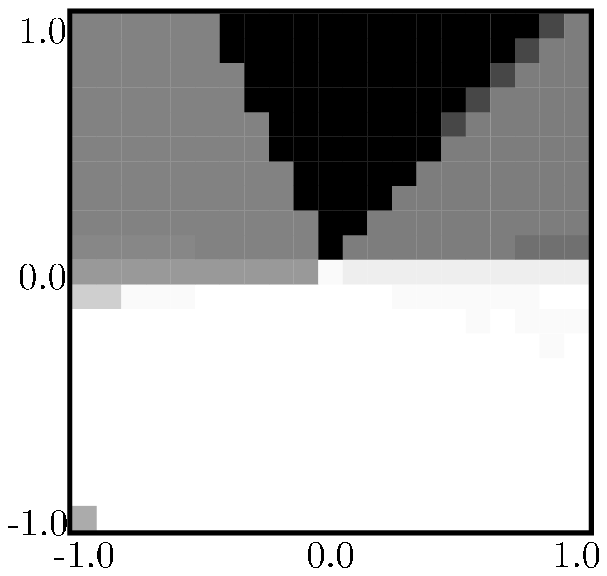}
}
\subfigure[Arcus1, $m=12$]{
\label{fig:tuning:c}
 \includegraphics[width=6cm]{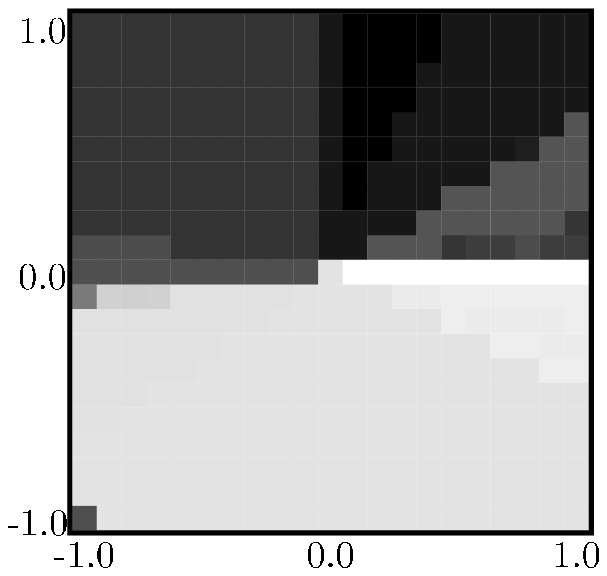}
}
\subfigure[Scholl, $m=38$]{
\label{fig:tuning:d}
 \includegraphics[width=6cm]{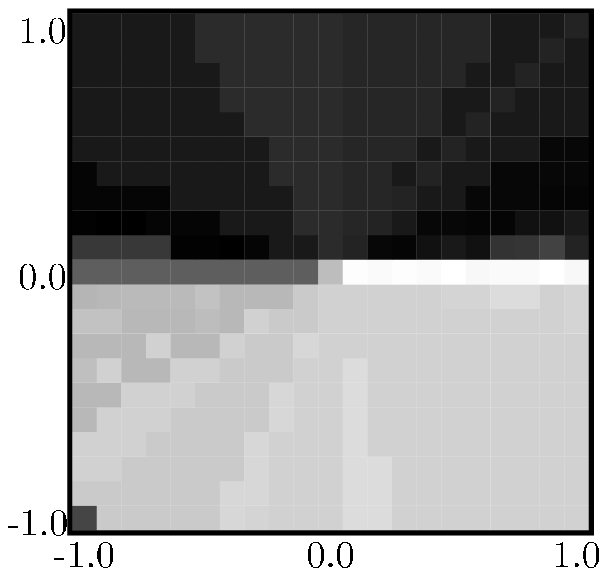}
}
\subfigure[Lutz1, $m=9$]{
\label{fig:tuning:e}
 \includegraphics[width=6cm]{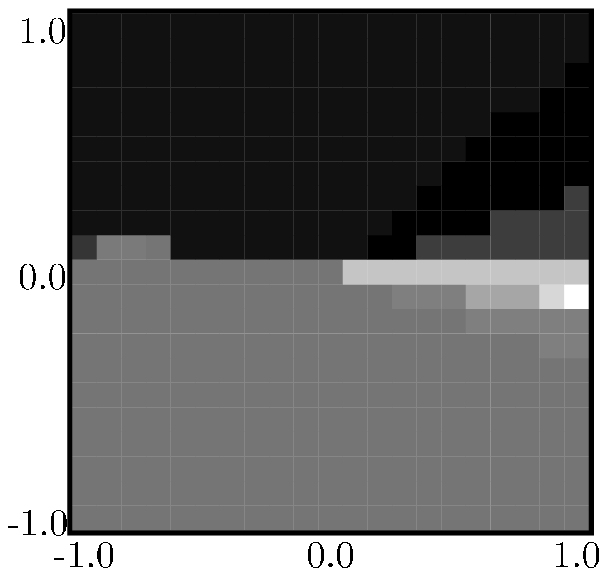}
}
\subfigure[Lutz1, $m=10$]{
\label{fig:tuning:f}
 \includegraphics[width=6cm]{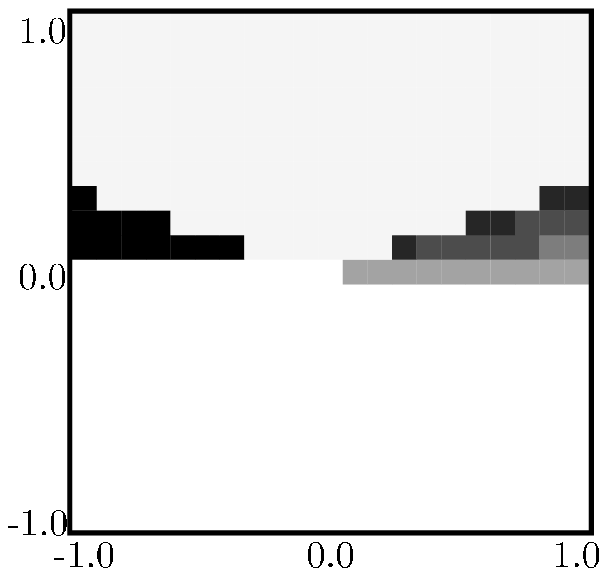}
}
\caption{Tuning results presented in graphical form for six representative instances. The y-axis ranges over the values of $\kappa_1$, while the x-axis ranges over the values of $\kappa_2$. The gray levels of the squares indicate the quality of the algorithm when run with the corresponding setting: the lighter a square is painted, the better is the parameter setting.}
\label{fig:tuning}
\end{figure}

In Figure~\ref{fig:tuning} tuning information is provided in graphical form for a few representative examples. The y-axis of the presented graphics varies over the different values of $\kappa_1$, while the x-axis ranges over the allowed values of $\kappa_2$. Note that each graphic consists of 441 squares representing the 441 different combinations of values for $\kappa_1$ and $\kappa_2$. The gray level in which each square is painted indicates the quality of the algorithm when run with the corresponding parameter setting. In particular, black color denotes the worst setting, whereas white color indicates the best algorithm setting. In some cases such as (Buxey, $m=7$) and (Tonge, $m=16$), as shown in Figures~\ref{fig:tuning:a} and~\ref{fig:tuning:b}, there is a wide range of good settings, which are basically all those with $\kappa_1 \leq 0$. In other examples such as (Arcus1, $m=12$) and (Scholl, $m=38$) it is strictly required to set $\kappa_1$ to 0 and $\kappa_2$ to a positive value for obtaining good solutions; see Figures~\ref{fig:tuning:c} and~\ref{fig:tuning:d}. Finally, the graphics shown in Figures~\ref{fig:tuning:e} and~\ref{fig:tuning:f} indicate that even for the same precedence graph a good parameter setting might depend strongly on the number of work stations. 

\begin{table}[!h]
\caption{\label{tab:tuning:q} Differences in algorithm performance when considering the best and the worst parameter setting.}
\begin{center}
\begin{tabular}{l|lll} \hline
{\bf Instance}  & {\bf Best setting} & {\bf Worst setting} & {\bf Difference (\%)} \\ \hline
(Buxey, $m=7$)   & 47                 & 52                  & 10.64 \\
(Tonge, $m=16$)  & 222                & 265                 & 19.37 \\
(Arcus1, $m=12$) & 12599              & 13767               & 9.27  \\
(Scholl, $m=38$) & 1857               & 2031                & 9.37  \\
(Lutz1, $m=9$)   & 1637               & 1801                & 10.02 \\
(Lutz1, $m=10$)  & 1525               & 1619                & 6.16  \\ \hline
\end{tabular}						       
\end{center}
\end{table}

It is also interesting to quantify the differences in algorithm performance for different parameter settings. Table~\ref{tab:tuning:q} shows for the six cases presented in Figure~\ref{fig:tuning} the result of the algorithm with the best setting (column {\bf Best setting}), the result of the algorithm with the worst setting (column {\bf Worst setting}), and the difference (in percent) between these two settings. The results in Table~\ref{tab:tuning:q} show that there are considerable differences in performance between the best and the worst algorithm setting. This underlines the importance of finding opportune values for $\kappa_1$ and $\kappa_2$.

\subsection{Results}

Algorithm \textsc{Ibs} was applied 20 times to all 302 instances. Herefore we used a computation time limit of 180 seconds for each cycle time, that is, \textsc{Ibs} was given maximally 180 seconds for finding a feasible solution for a given cycle time. In case of success, the algorithm has again 180 seconds for the next smaller cycle time, etc. Detailed results of \textsc{Ibs} for all 302 instances are given in Tables~\ref{tab:results:A} and~\ref{tab:results:B} that are to be found in Appendix~A. The data is, in both tables, presented as follows. The first two columns provide the name of the precedence graph and the number of work stations ($m$). The third column provides the value of the optimal (respectively, best-known) solutions. In case a value is not proved to be optimal it is overlined. More in detail, in 15 out of 302 cases optimality has not been proved yet. The remaining five columns are reserved for the results of \textsc{Ibs}. The first of these five columns contains the value of the best solution found by \textsc{Ibs} over 20 runs. In case this value is presented with a gray background, a new best-known solution has been found. On the other side, if this value is marked by an asterisk, the obtained result does not reach the value of a best-known solution. In all other cases the values correspond to the values of best-known solutions. The second column provides the average over 20 runs, while the third column contains the corresponding standard deviation. The fourth column gives the average time (in seconds) at which the best solution of a run was found, averaged over 20 runs. The fifth column provides the corresponding standard deviation. From the results presented in Tables~\ref{tab:results:A} and~\ref{tab:results:B} (see Appendix~A) we can observe that \textsc{Ibs} obtains best-known (respectively, optimal) solutions in 276 out of 302 cases. Moreover, new best solutions are obtained in 6 cases. This is remarkable as---despite a considerable amount of ongoing research---in the last 14 years no improved solutions have been reported. Only in 20 cases (all concerning precedence graphs Arcus1, Arcus2, Scholl, and Warnecke) our algorithm was not able to find the best solutions known. However, in most of these cases the deviation from the best-known solution is no more than one unit of cycle time.

\begin{table}[!h]
\caption{\label{tab:summary:total} Results of \textsc{Ibs} in comparison to the best (\textsc{TabuSearch}), respectively most recent (\textsc{DE\_rks} and \textsc{PNA-for}), methods from the literature.}
\begin{center}
\begin{tabular}{l|llll} \hline
         & \textsc{DE\_rks} & \textsc{PNA-for} & \textsc{TabuSearch} & {\bf Iterative Beam Search} (\textsc{Ibs}) \\ \hline
{\bf \#opt}    & n/g           & 39            & 168               & 282     \\
{\bf mrd (\%)} & 2.64          & 2.73          & 0.40              & 0.0029  \\
{\bf time (s)} & 10.74         & 407.28        & 84.90             & 31.61   \\ \hline
\multicolumn{4}{l}{\tiny Note: n/g means \textbf{not given}}
\end{tabular}						       
\end{center}
\end{table}

In addition to Tables~\ref{tab:results:A} and~\ref{tab:results:B} the results of \textsc{Ibs} are presented in a summarized way in Table~\ref{tab:summary:total}, in comparison to three other algorithms. \textsc{TabuSearch}~\cite{SchVos96:ts}, even though already published in 1996, still counts as the current state-of-the-art heuristic method for SALBP-2. \textsc{DE\_rks} is the best version of a differential evolution (DE) algorithm proposed in~\cite{Nea07:ijamt}, and \textsc{PNA-for} is a Petri net-based heuristic published in~\cite{Kil10:ijamt}. The last two methods are, to our knowledge, the most recently published heuristic methods for SALBP-2. Three measure are used in Table~\ref{tab:summary:total} for the comparison of \textsc{Ibs} with these three algorithms. The row labelled {\bf \#opt} provides the number of best-known solutions found by each method (over 302). Moreover, the row labelled {\bf mrd (\%)} gives the \emph{mean relative deviation (in percent)} of the results obtained by the four algorithms from the best-known solutions for all 302 instances. Finally, row {\bf time} contains the average computation time of the algorithms for all 302 instances. Concerning the quality of the results, we can conclude that \textsc{Ibs} clearly outperforms its competitors. For the correct interpretation of the computation times it has to be taken into account that the four algorithms were executed on computers with very different processor speeds. While \textsc{TabuSearch} was executed on a 80486 DX2-66 processor, \textsc{PNA-for} was run on an Athlon XP 2000+ processor with 1.67 GHz, and \textsc{DE\_rks} was run on a Pentium IV processor with 1.7 GHz. This means that \textsc{TabuSearch} was run by far on the slowest machine, \textsc{Ibs} by far on the fastest machine, and \textsc{PNA-for} and \textsc{DE\_rks} on comparable machines. Given the computation times in Table~\ref{tab:summary:total} we can savely conclude that \textsc{TabuSearch} is the fastest algorithm, and \textsc{PNA-for} is the slowest one. However, note that assembly line balancing is, in most cases, not a time-critical application. In other words, for most practical purposes it does not matter if an algorithm takes 1 minute or 6 hours of computation time. 

In the following we present the results of \textsc{Ibs} in comparison to \textsc{DE\_rks} and \textsc{PNA-for} in the same way as done in~\cite{Nea07:ijamt} and~\cite{Kil10:ijamt}. In these works, results were presented as averages over instances based on the same precedence graph, and also averaged over {\sf Dataset1} and {\sf Dataset2}. The quality of the results is given in terms of the \emph{mean relative deviation (in percent)} from the best-known solutions. Tables~\ref{tab:summary:set1} and~\ref{tab:summary:set2} clearly show that \textsc{Ibs} is largely superior to both competitor algorithms.

\begin{table}[t]
\caption{\label{tab:summary:set1} Results of \textsc{Ibs} in comparison to \textsc{DE\_rks} and \textsc{PNA-for} for the 128 instances of {\sf Dataset1} (averaged over precedence graphs).}
\begin{center}
\begin{tabular}{l|ll|l|ll} \hline
{\bf Graph} & \multicolumn{2}{c|}{\textsc{DE\_rks}} & {\textsc{PNA-for}} & \multicolumn{2}{c}{{\bf Iterative Beam Search} (\textsc{Ibs})} \\
         & mrd (\%) & time (s) & mrd (\%) & mrd (\%) & time (s) \\ \hline
Buxey          & 1.16       & 0.80       & 3.07       & 0.0          & 0.06   \\
Sawyer         & 2.27       & 1.64       & 4.00       & 0.0          & 0.14   \\
Lutz1          & 0.32       & 0.88       & 4.09       & 0.0          & 0.55   \\
Gunther        & 0.14       & 1.08       & 1.27       & 0.0          & 0.02   \\
Kilbridge      & 0.66       & 1.43       & 2.19       & 0.0          & 0.0067 \\
Tonge          & 1.88       & 3.71       & 2.53       & 0.0          & 7.73   \\
Arcus1         & 0.99       & 5.29       & 2.47       & 0.0287       & 152.34 \\
Lutz2          & 3.08       & 1.00       & 2.99       & 0.0          & 0.035  \\
Arcus2         & 4.96       & 19.02      & 2.06       & 0.0066       & 135.83 \\ \hline
{\bf Average:} & {\bf 1.72} & {\bf 3.87} & {\bf 2.57} & {\bf 0.0058} & {\bf 51.76} \\ \hline
\end{tabular}
\end{center}
\end{table}

\begin{table}[t]
\caption{\label{tab:summary:set2} Results of \textsc{Ibs} in comparison to \textsc{DE\_rks} and \textsc{PNA-for} for the 174 instances of {\sf Dataset2} (averaged over precedence graphs).}
\begin{center}
\begin{tabular}{l|ll|l|ll} \hline
{\bf Graph} & \multicolumn{2}{c|}{\textsc{DE\_rks}} & {\textsc{PNA-for}} & \multicolumn{2}{c}{{\bf Iterative Beam Search} (\textsc{Ibs})} \\
         & mrd (\%) & time (s) & mrd (\%) & mrd (\%) & time (s) \\ \hline
Hahn           & 0.0        & 1.00        & 2.52       & 0.0          & 0.065  \\
Warnecke       & 3.74       & 3.53        & 5.57       & 0.0579       & 1.54   \\
Wee-Mag        & 1.23       & 3.68        & 1.56       & 0.0          & 0.65   \\
Lutz3          & 1.68       & 5.62        & 2.59       & 0.0          & 0.61   \\
Mukherje       & n/a        & n/a         & 1.04       & 0.0          & 2.52   \\
Barthold       & 0.26       & 19.47       & 1.02       & 0.0          & 0.079  \\
Barthol2       & 6.85       & 33.08       & 3.97       & 0.0          & 0.96   \\
Scholl         & 9.51       & 43.95       & 3.21       & 0.0028       & 98.69  \\ \hline
{\bf Average:} & {\bf 3.32} & {\bf 15.79} & {\bf 2.85} & {\bf 0.00071} & {\bf 16.79} \\ \hline
\end{tabular}						       
\end{center}
\end{table}

\subsection{Results of a High-Performance Version}

In an attempt to further improve on the results of our algorithm we decided to apply a high-performance version of \textsc{Ibs} to all problem instances for which the optimal solution is unknown and, additionally, to all instances where \textsc{Ibs}---with the settings as outlined in the previous section---was not able to find a best-known solution. This high-performance version is obtained as follows. First, 1800 seconds are used as a time limit for each cycle time. Second, in line 17 of Algorithm~\ref{algo:scheme} only 1\% of the time limit is used (instead of 5\%). Third, for each application of beam search in lines 18 and 20 of Algorithm~\ref{algo:scheme} the beam width $\kbw$ is randomly chosen from $[150,250]$ and the number of extensions is randomly chosen from $[20,40]$. Moreover, with a probability of $0.5$ the heuristic information is---for each application of beam search---calculated using the weight values as outlined in Table~\ref{tab:tuning}. Otherwise, the weight values are chosen randomly from $[-1,1]$. With these modifications we applied \textsc{Ibs} exactly once to all the instances of Table~\ref{tab:high-performance}. The results of the algorithm are given in column {\bf Result}. Indeed, the number of instances for which a best-known solution can not be found before is reduced from 20 to 10 instances. Moreover, the algorithm is now able to find new best-known solutions in 9 (instead of only 6) cases. Summarizing, this amounts to 283 best-known solutions found and 9 new best-known solutions obtained. In one case, (Scholl, $m = 49$), the new best-known solution is provenly optimal, as its value coincides with the best known lower bound. 


\begin{table}[t]
\caption{\label{tab:high-performance} Results of a high-performance version of \textsc{Ibs}.}
\begin{center}
\begin{tabular}{ll|l||ll|l} \hline
{\bf Graph} & $m$ & {\bf Result} & {\bf Graph} & $m$ & {\bf Result} \\ \hline
Arcus1 (83)   & 8  & 9554  & Arcus2 (111)  & 23 & \colorbox{lightgray}{6560} \\
              & 11 & 7085$^{\ast}$  & 		   & 24 & \colorbox{lightgray}{6282} \\
              & 12 & 6412  & 		   & 25 & \colorbox{lightgray}{6101} \\
              & 17 & 4527$^{\ast}$  & 		   & 26 & \colorbox{lightgray}{5855} \\ \cline{4-6}
              & 18 & 4323$^{\ast}$  & Scholl (297)  & 31 & 2247 \\
              & 19 & 4071$^{\ast}$  & 		   & 36 & 1936$^{\ast}$ \\
              & 20 & 3886$^{\ast}$  & 		   & 42 & 1660$^{\ast}$ \\ \cline{1-3}
Arcus2 (111)  & 14 & 10747 & 		   & 46 & 1515 \\
              & 15 & 10035 & 		   & 47 & 1484$^{\ast}$ \\
              & 16 & 9413$^{\ast}$  & 		   & 49 & \colorbox{lightgray}{1423} \\ \cline{4-6}
              & 17 & 8857$^{\ast}$  & Warnecke (58) & 25 & 64   \\ \cline{4-6}
              & 18 & 8377  & Wee-Mag (75)  & 18 & 87   \\
              & 19 & \colorbox{lightgray}{7922}  & 		   & 19 & 85   \\
              & 20 & \colorbox{lightgray}{7524}  & 		   & 23 & 67   \\
              & 21 & \colorbox{lightgray}{7187}  & 		   & 27 & 65   \\
              & 22 & \colorbox{lightgray}{6856}  & 		   & 28 & 64   \\ \hline
\end{tabular}						       
\end{center}
\end{table}

\section{Conclusions and Future Work}
\label{sec:conclusions}

In this work we have proposed an iterative beam search algorithm for the simple assembly line balancing problem with a fixed number of work stations, SALBP-2. The experimental evalution of the algorithm has shown that it is currently a state-of-the-art method for this problem. Appart from producing optimal, respectively best-known, solutions in 283 out of 302 test cases, our algorithm generated new best-known solutions in further 9 test cases. Encouraged by the results for the SALBP-1 version of the problem (as published in~\cite{Blu08:informs}) and the results obtained in this paper for the SALBP-2 we intent to apply similar algorithms based on beam search to other assembly line balancing problems.

\section*{Acknowledgements}

This work was supported by grant TIN2007-66523 (FORMALISM) of the Spanish government. Moreover, Christian Blum acknowledges support from the \textit{Ram\'on y Cajal} program of the Spanish Ministry of Science and Innovation.

Many thanks go to Armin Scholl for verifying the new best-known solutions found by the algorithm proposed in this work. Finally, we would also like to express our thanks to Crist{\'o}bal Miralles who was involved as a co-author of a similar work for the more general assembly line worker assignment and balancing problem.

\bibliographystyle{plain}

\pagebreak

\begin{appendix}

\section*{Appendix A}

\begin{table}[!b]
\caption{\label{tab:results:A} Detailed results of \textsc{Ibs} for 302 test instances (Part A).}
\begin{center}
\scalebox{0.58}{
\begin{tabular}{ll|l|lllll||ll|l|lllll} \hline
Graph & $m$ & {\bf bks} & \multicolumn{5}{|c||}{{\bf Iterative Beam Search} (\textsc{Ibs})} & Graph & $m$ & {\bf bks} & \multicolumn{5}{|c}{{\bf Iterative Beam Search} (\textsc{Ibs})} \\
         &     &            & {\bf best} & {\bf avg} & {\bf std} & {\bf time (s)} & {\bf std} & &     &            & {\bf best} & {\bf avg} & {\bf std} & {\bf time (s)} & {\bf std} \\ \hline
Arcus1 (83) & 3 & 25236 & 25236 & 25236.00 & (0.00) & 0.56 & (0.29)                  &  Barthold (148) & 9 & 626 & 626 & 626.00 & (0.00) & 0.13 & (0.03) \\
 & 4 & 18927 & 18927 & 18927.00 & (0.00) & 6.16 & (3.92) 			&   & 10 & 564 & 564 & 564.00 & (0.00) & 0.10 & (0.02) \\
 & 5 & 15142 & 15142 & 15142.00 & (0.00) & 80.08 & (16.69) 			&   & 11 & 513 & 513 & 513.00 & (0.00) & 0.09 & (0.00) \\
 & 6 & 12620 & 12620 & 12620.00 & (0.00) & 15.52 & (5.76) 			&   & 12 & 470 & 470 & 470.00 & (0.00) & 0.09 & (0.00) \\
 & 7 & 10826 & 10826 & 10826.00 & (0.00) & 59.80 & (19.89) 			&   & 13 & 434 & 434 & 434.00 & (0.00) & 0.16 & (0.02) \\
 & 8 & 9554 & $^{\ast}$9555 & 9555.55 & (0.51) & 120.54 & (63.16) 		&   & 14 & 403 & 403 & 403.00 & (0.00) & 0.14 & (0.01) \\
 & 9 & 8499 & 8499 & 8500.70 & (0.80) & 130.73 & (117.23) 			&   & 15 & 383 & 383 & 383.00 & (0.00) & 0.05 & (0.00) \\ \cline{9-16}
 & 10 & 7580 & 7580 & 7580.95 & (0.60) & 311.89 & (109.87) 			&  Buxey (29) & 7 & 47 & 47 & 47.00 & (0.00) & 0.46 & (0.51) \\
 & 11 & 7084 & $^{\ast}$7086 & 7086.70 & (0.47) & 71.13 & (52.77) 		&   & 8 & 41 & 41 & 41.00 & (0.00) & 0.00 & (0.00) \\
 & 12 & 6412 & $^{\ast}$6413 & 6414.10 & (0.72) & 362.32 & (66.26) 		&   & 9 & 37 & 37 & 37.00 & (0.00) & 0.01 & (0.00) \\
 & 13 & 5864 & 5864 & 5864.00 & (0.00) & 129.18 & (12.99) 			&   & 10 & 34 & 34 & 34.00 & (0.00) & 0.00 & (0.00) \\
 & 14 & 5441 & 5441 & 5441.00 & (0.00) & 2.02 & (0.41)  			&   & 11 & 32 & 32 & 32.00 & (0.00) & 0.00 & (0.00) \\
 & 15 & 5104 & 5104 & 5104.45 & (0.60) & 222.29 & (73.92) 			&   & 12 & 28 & 28 & 28.00 & (0.00) & 0.01 & (0.00) \\
 & 16 & 4850 & 4850 & 4850.00 & (0.00) & 33.84 & (9.59) 			&   & 13 & 27 & 27 & 27.00 & (0.00) & 0.01 & (0.00) \\
 & 17 & 4516 & $^{\ast}$4524 & 4526.20 & (0.89) & 282.49 & (100.96) 		&   & 14 & 25 & 25 & 25.00 & (0.00) & 0.00 & (0.00) \\ \cline{9-16}
 & 18 & 4317 & $^{\ast}$4322 & 4323.20 & (0.77) & 291.73 & (111.23) 		&  Gunther (35) & 6 & 84 & 84 & 84.00 & (0.00) & 0.00 & (0.00) \\
 & 19 & 4068 & $^{\ast}$4073 & 4074.80 & (1.01) & 326.78 & (125.49) 		&   & 7 & 72 & 72 & 72.00 & (0.00) & 0.03 & (0.00) \\
 & 20 & 3882 & $^{\ast}$3886 & 3889.25 & (2.05) & 594.21 & (204.09) 		&   & 8 & 63 & 63 & 63.00 & (0.00) & 0.01 & (0.00) \\
 & 21 & 3691 & 3691 & 3691.00 & (0.00) & 5.58 & (0.81)  			&   & 9 & 54 & 54 & 54.00 & (0.00) & 0.05 & (0.04) \\
 & 22 & 3691 & 3691 & 3691.00 & (0.00) & 0.02 & (0.00)  			&   & 10 & 50 & 50 & 50.00 & (0.00) & 0.01 & (0.00) \\ \cline{1-8}
Arcus2 (111) & 3 & 50133 & 50133 & 50133.00 & (0.00) & 0.09 & (0.03)  		&   & 11 & 48 & 48 & 48.00 & (0.00) & 0.00 & (0.00) \\
 & 4 & 37600 & 37600 & 37600.00 & (0.00) & 0.41 & (0.02) 			&   & 12 & 44 & 44 & 44.00 & (0.00) & 0.01 & (0.00) \\
 & 5 & 30080 & 30080 & 30080.00 & (0.00) & 0.43 & (0.22) 			&   & 13 & 42 & 42 & 42.00 & (0.00) & 0.01 & (0.00) \\
 & 6 & 25067 & 25067 & 25067.00 & (0.00) & 1.13 & (0.59) 			&   & 14 & 40 & 40 & 40.00 & (0.00) & 0.02 & (0.00) \\
 & 7 & 21486 & 21486 & 21486.00 & (0.00) & 1.76 & (1.04) 			&   & 15 & 40 & 40 & 40.00 & (0.00) & 0.01 & (0.00) \\ \cline{9-16}
 & 8 & 18800 & 18800 & 18800.00 & (0.00) & 11.36 & (2.86) 			&  Hahn (53) & 3 & 4787 & 4787 & 4787.00 & (0.00) & 0.00 & (0.00) \\
 & 9 & 16711 & 16711 & 16711.00 & (0.00) & 28.03 & (12.51) 			&   & 4 & 3677 & 3677 & 3677.00 & (0.00) & 0.08 & (0.01) \\
 & 10 & 15040 & 15040 & 15040.00 & (0.00) & 38.67 & (15.08) 			&   & 5 & 2823 & 2823 & 2823.00 & (0.00) & 0.01 & (0.00) \\
 & 11 & 13673 & 13673 & 13673.00 & (0.00) & 49.47 & (12.83) 			&   & 6 & 2400 & 2400 & 2400.00 & (0.00) & 0.05 & (0.01) \\
 & 12 & 12534 & 12534 & 12534.00 & (0.00) & 43.34 & (15.17) 			&   & 7 & 2336 & 2336 & 2336.00 & (0.00) & 0.29 & (0.01) \\
 & 13 & 11570 & 11570 & 11570.00 & (0.00) & 32.60 & (14.63) 			&   & 8 & 1907 & 1907 & 1907.00 & (0.00) & 0.09 & (0.03) \\
 & 14 & 10747 & $^{\ast}$10748 & 10748.00 & (0.00) & 62.69 & (16.17) 		&   & 9 & 1827 & 1827 & 1827.00 & (0.00) & 0.00 & (0.00) \\
 & 15 & 10035 & $^{\ast}$10036 & 10036.40 & (0.50) & 112.54 & (63.37) 		&   & 10 & 1775 & 1775 & 1775.00 & (0.00) & 0.00 & (0.00) \\ \cline{9-16}
 & 16 & 9412 & $^{\ast}$9416 & 9416.60 & (0.68) & 290.35 & (90.83) 		&  Kilbridge (45) & 3 & 184 & 184 & 184.00 & (0.00) & 0.00 & (0.00) \\
 & 17 & 8855 & $^{\ast}$8864 & 8864.90 & (0.31) & 87.41 & (52.86) 		&   & 4 & 138 & 138 & 138.00 & (0.00) & 0.01 & (0.00) \\
 & 18 & $\overline{8377}$ & 8377 & 8377.00 & (0.00) & 8.11 & (1.00)  			&   & 5 & 111 & 111 & 111.00 & (0.00) & 0.00 & (0.00) \\
 & 19 & $\overline{7928}$ & \colorbox{lightgray}{7924} & 7925.60 & (0.60) & 205.09 & (76.12) &   & 6 & 92 & 92 & 92.00 & (0.00) & 0.00 & (0.00) \\
 & 20 & $\overline{7526}$ & \colorbox{lightgray}{7524} & 7524.40 & (0.50) & 159.72 & (57.92) &   & 7 & 79 & 79 & 79.00 & (0.00) & 0.01 & (0.00) \\
 & 21 & $\overline{7188}$ & $^{\ast}$7192 & 7193.40 & (0.82) & 318.78 & (89.64) 		&   & 8 & 69 & 69 & 69.00 & (0.00) & 0.01 & (0.00) \\
 & 22 & $\overline{6859}$ & \colorbox{lightgray}{6858} & 6858.20 & (0.41) & 226.94 & (75.52) &   & 9 & 62 & 62 & 62.00 & (0.00) & 0.01 & (0.00) \\
 & 23 & $\overline{6561}$ & \colorbox{lightgray}{6560} & 6563.10 & (1.45) & 428.68 & (168.53)&   & 10 & 56 & 56 & 56.00 & (0.00) & 0.01 & (0.00) \\
 & 24 & $\overline{6289}$ & \colorbox{lightgray}{6284} & 6285.65 & (1.46) & 311.77 & (144.55)&   & 11 & 55 & 55 & 55.00 & (0.00) & 0.01 & (0.00) \\ \cline{9-16}
 & 25 & $\overline{6106}$ & $^{\ast}$6112 & 6114.15 & (0.99) & 305.50 & (90.99) 		&  Lutz1 (32) & 8 & 1860 & 1860 & 1860.00 & (0.00) & 0.14 & (0.00) \\
 & 26 & $\overline{5856}$ & $^{\ast}$5858 & 5860.45 & (1.76) & 661.48 & (192.47) 		&   & 9 & 1638 & 1638 & 1638.00 & (0.00) & 2.41 & (0.34) \\
 & 27 & 5689 & 5689 & 5689.00 & (0.00) & 9.29 & (0.17)  			&   & 10 & 1526 & 1526 & 1526.00 & (0.00) & 0.18 & (0.01) \\ \cline{1-8}
Barthol2 (148) & 27 & 157 & 157 & 157.00 & (0.00) & 0.31 & (0.03) 			&   & 11 & 1400 & 1400 & 1400.00 & (0.00) & 0.01 & (0.00) \\
 & 28 & 152 & 152 & 152.00 & (0.00) & 0.44 & (0.07) 				&   & 12 & 1400 & 1400 & 1400.00 & (0.00) & 0.00 & (0.00) \\ \cline{9-16}
 & 29 & 146 & 146 & 146.00 & (0.00) & 0.80 & (0.32) 				&  Lutz2 (89) & 9 & 54 & 54 & 54.00 & (0.00) & 0.02 & (0.01) \\
 & 30 & 142 & 142 & 142.00 & (0.00) & 0.15 & (0.03) 				&   & 10 & 49 & 49 & 49.00 & (0.00) & 0.03 & (0.00) \\
 & 31 & 137 & 137 & 137.00 & (0.00) & 0.67 & (0.09) 				&   & 11 & 45 & 45 & 45.00 & (0.00) & 0.04 & (0.00) \\
 & 32 & 133 & 133 & 133.00 & (0.00) & 0.57 & (0.07) 				&   & 12 & 41 & 41 & 41.00 & (0.00) & 0.03 & (0.00) \\
 & 33 & 129 & 129 & 129.00 & (0.00) & 0.52 & (0.06) 				&   & 13 & 38 & 38 & 38.00 & (0.00) & 0.03 & (0.00) \\
 & 34 & 125 & 125 & 125.00 & (0.00) & 0.47 & (0.05) 				&   & 14 & 35 & 35 & 35.00 & (0.00) & 0.03 & (0.00) \\
 & 35 & 121 & 121 & 121.00 & (0.00) & 0.94 & (0.15) 				&   & 15 & 33 & 33 & 33.00 & (0.00) & 0.03 & (0.00) \\
 & 36 & 118 & 118 & 118.00 & (0.00) & 0.87 & (0.11) 				&   & 16 & 31 & 31 & 31.00 & (0.00) & 0.02 & (0.00) \\
 & 37 & 115 & 115 & 115.00 & (0.00) & 0.92 & (0.19) 				&   & 17 & 29 & 29 & 29.00 & (0.00) & 0.04 & (0.00) \\
 & 38 & 112 & 112 & 112.00 & (0.00) & 0.76 & (0.06) 				&   & 18 & 28 & 28 & 28.00 & (0.00) & 0.04 & (0.00) \\
 & 39 & 109 & 109 & 109.00 & (0.00) & 0.70 & (0.13) 				&   & 19 & 26 & 26 & 26.00 & (0.00) & 0.04 & (0.01) \\
 & 40 & 106 & 106 & 106.00 & (0.00) & 0.91 & (0.32) 				&   & 20 & 25 & 25 & 25.00 & (0.00) & 0.04 & (0.00) \\
 & 41 & 104 & 104 & 104.00 & (0.00) & 0.72 & (0.05) 				&   & 21 & 24 & 24 & 24.00 & (0.00) & 0.02 & (0.00) \\
 & 42 & 101 & 101 & 101.00 & (0.00) & 2.77 & (5.43) 				&   & 22 & 23 & 23 & 23.00 & (0.00) & 0.04 & (0.00) \\
 & 43 & 99 & 99 & 99.00 & (0.00) & 0.84 & (0.23) 				&   & 23 & 22 & 22 & 22.00 & (0.00) & 0.05 & (0.00) \\
 & 44 & 97 & 97 & 97.00 & (0.00) & 0.59 & (0.15) 				&   & 24 & 21 & 21 & 21.00 & (0.00) & 0.05 & (0.00) \\
 & 45 & 95 & 95 & 95.00 & (0.00) & 0.76 & (0.25) 				&   & 25 & 20 & 20 & 20.00 & (0.00) & 0.02 & (0.00) \\
 & 46 & 93 & 93 & 93.00 & (0.00) & 0.68 & (0.29) 				&   & 26 & 19 & 19 & 19.00 & (0.00) & 0.05 & (0.02) \\
 & 47 & 91 & 91 & 91.00 & (0.00) & 0.67 & (0.15) 				&   & 27 & 19 & 19 & 19.00 & (0.00) & 0.03 & (0.00) \\
 & 48 & 89 & 89 & 89.00 & (0.00) & 1.17 & (0.47) 				&   & 28 & 18 & 18 & 18.00 & (0.00) & 0.05 & (0.00) \\ \cline{9-16}
 & 49 & 87 & 87 & 87.00 & (0.00) & 0.83 & (0.26) 				&  Lutz3 (89) & 3 & 548 & 548 & 548.00 & (0.00) & 0.01 & (0.00) \\
 & 50 & 85 & 85 & 85.90 & (0.31) & 3.28 & (10.33) 				&   & 4 & 411 & 411 & 411.00 & (0.00) & 0.03 & (0.00) \\
 & 51 & 84 & 84 & 84.00 & (0.00) & 2.60 & (2.09)        			&   & 5 & 329 & 329 & 329.00 & (0.00) & 0.01 & (0.00) \\ \cline{1-8}
Barthold (148) & 3 & 1878 & 1878 & 1878.00 & (0.00) & 0.01 & (0.00) 			&   & 6 & 275 & 275 & 275.00 & (0.00) & 0.01 & (0.00) \\
 & 4 & 1409 & 1409 & 1409.00 & (0.00) & 0.02 & (0.00) 				&   & 7 & 236 & 236 & 236.00 & (0.00) & 0.02 & (0.00) \\
 & 5 & 1127 & 1127 & 1127.00 & (0.00) & 0.02 & (0.00) 				&   & 8 & 207 & 207 & 207.00 & (0.00) & 0.04 & (0.00) \\
 & 6 & 939 & 939 & 939.00 & (0.00) & 0.06 & (0.01) 				&   & 9 & 184 & 184 & 184.00 & (0.00) & 5.54 & (4.14) \\
 & 7 & 805 & 805 & 805.00 & (0.00) & 0.11 & (0.00) 				&   & 10 & 165 & 165 & 165.00 & (0.00) & 0.09 & (0.01) \\
 & 8 & 705 & 705 & 705.00 & (0.00) & 0.04 & (0.01)                              &   & 11 & 151 & 151 & 151.00 & (0.00) & 0.06 & (0.00) \\ \hline
\end{tabular}}
\end{center}
\end{table}

\begin{table}[p]
\caption{\label{tab:results:B} Detailed results of \textsc{Ibs} for 302 test instances (Part B).}
\begin{center}
\scalebox{0.58}{
\begin{tabular}{ll|l|lllll||ll|l|lllll} \hline
Graph & $m$ & {\bf bks} & \multicolumn{5}{|c||}{{\bf Iterative Beam Search} (\textsc{Ibs})} & Graph & $m$ & {\bf bks} & \multicolumn{5}{|c}{{\bf Iterative Beam Search} (\textsc{Ibs})} \\
         &     &            & {\bf best} & {\bf avg} & {\bf std} & {\bf time (s)} & {\bf std} & &     &            & {\bf best} & {\bf avg} & {\bf std} & {\bf time (s)} & {\bf std} \\ \hline
Lutz3 (89) & 12 & 138 & 138 & 138.00 & (0.00) & 0.02 & (0.00)                          & Tonge (70) & 6 & 585 & 585 & 585.00 & (0.00) & 0.03 & (0.02) \\
 & 13 & 128 & 128 & 128.00 & (0.00) & 0.05 & (0.03) 				    &  & 7 & 502 & 502 & 502.00 & (0.00) & 0.06 & (0.01) \\
 & 14 & 118 & 118 & 118.00 & (0.00) & 0.67 & (0.52) 				    &  & 8 & 439 & 439 & 439.00 & (0.00) & 0.04 & (0.01) \\
 & 15 & 110 & 110 & 110.00 & (0.00) & 0.80 & (0.66) 				    &  & 9 & 391 & 391 & 391.00 & (0.00) & 0.05 & (0.00) \\
 & 16 & 105 & 105 & 105.00 & (0.00) & 0.34 & (0.30) 				    &  & 10 & 352 & 352 & 352.00 & (0.00) & 0.06 & (0.01) \\
 & 17 & 98 & 98 & 98.00 & (0.00) & 0.19 & (0.05) 				    &  & 11 & 320 & 320 & 320.00 & (0.00) & 0.06 & (0.02) \\
 & 18 & 93 & 93 & 93.00 & (0.00) & 0.16 & (0.05) 				    &  & 12 & 294 & 294 & 294.00 & (0.00) & 0.09 & (0.01) \\
 & 19 & 89 & 89 & 89.00 & (0.00) & 0.05 & (0.03) 				    &  & 13 & 271 & 271 & 271.00 & (0.00) & 0.06 & (0.01) \\
 & 20 & 85 & 85 & 85.00 & (0.00) & 0.14 & (0.05) 				    &  & 14 & 251 & 251 & 251.80 & (0.41) & 20.25 & (45.26) \\
 & 21 & 80 & 80 & 80.00 & (0.00) & 0.11 & (0.02) 				    &  & 15 & 235 & 235 & 235.00 & (0.00) & 0.35 & (0.16) \\
 & 22 & 76 & 76 & 76.00 & (0.00) & 4.20 & (3.30) 				    &  & 16 & 221 & 221 & 221.00 & (0.00) & 0.97 & (0.93) \\
 & 23 & 74 & 74 & 74.00 & (0.00) & 0.29 & (0.18)  			    &  & 17 & 208 & 208 & 208.00 & (0.00) & 1.02 & (0.95) \\ \cline{1-8}
Mukherje (94) & 3 & 1403 & 1403 & 1403.00 & (0.00) & 0.01 & (0.00) 		    &  & 18 & 196 & 196 & 196.15 & (0.37) & 46.76 & (39.05) \\
 & 4 & 1052 & 1052 & 1052.00 & (0.00) & 0.10 & (0.00) 			    &  & 19 & 186 & 186 & 186.35 & (0.49) & 87.09 & (64.62) \\
 & 5 & 844 & 844 & 844.00 & (0.00) & 0.02 & (0.00) 				    &  & 20 & 177 & 177 & 177.90 & (0.31) & 9.46 & (39.39) \\
 & 6 & 704 & 704 & 704.00 & (0.00) & 0.02 & (0.00) 				    &  & 21 & 170 & 170 & 170.00 & (0.00) & 8.71 & (9.43) \\
 & 7 & 621 & 621 & 621.00 & (0.00) & 0.12 & (0.00) 				    &  & 22 & 162 & 162 & 162.00 & (0.00) & 2.52 & (2.45) \\
 & 8 & 532 & 532 & 532.00 & (0.00) & 0.09 & (0.01) 				    &  & 23 & 156 & 156 & 156.00 & (0.00) & 0.11 & (0.04) \\
 & 9 & 471 & 471 & 471.00 & (0.00) & 0.09 & (0.01) 				    &  & 24 & 156 & 156 & 156.00 & (0.00) & 0.02 & (0.00) \\
 & 10 & 424 & 424 & 424.00 & (0.00) & 0.02 & (0.00) 				    &  & 25 & 156 & 156 & 156.00 & (0.00) & 0.02 & (0.00) \\ \cline{9-16}
 & 11 & 391 & 391 & 391.00 & (0.00) & 0.10 & (0.01) 				    & Warnecke (58) & 3 & 516 & 516 & 516.00 & (0.00) & 0.02 & (0.00) \\
 & 12 & 358 & 358 & 358.00 & (0.00) & 0.05 & (0.01) 				    &  & 4 & 387 & 387 & 387.00 & (0.00) & 0.08 & (0.08) \\
 & 13 & 325 & 325 & 325.00 & (0.00) & 0.17 & (0.06) 				    &  & 5 & 310 & 310 & 310.00 & (0.00) & 0.03 & (0.00) \\
 & 14 & 311 & 311 & 311.00 & (0.00) & 0.08 & (0.01) 				    &  & 6 & 258 & 258 & 258.00 & (0.00) & 0.10 & (0.06) \\
 & 15 & 288 & 288 & 288.00 & (0.00) & 0.09 & (0.02) 				    &  & 7 & 222 & 222 & 222.00 & (0.00) & 0.03 & (0.00) \\
 & 16 & 268 & 268 & 268.00 & (0.00) & 0.20 & (0.02) 				    &  & 8 & 194 & 194 & 194.00 & (0.00) & 0.05 & (0.01) \\
 & 17 & 251 & 251 & 251.00 & (0.00) & 0.45 & (0.05) 				    &  & 9 & 172 & 172 & 172.00 & (0.00) & 2.40 & (2.28) \\
 & 18 & 239 & 239 & 239.00 & (0.00) & 0.24 & (0.02) 				    &  & 10 & 155 & 155 & 155.00 & (0.00) & 0.05 & (0.01) \\
 & 19 & 226 & 226 & 226.00 & (0.00) & 0.03 & (0.01) 				    &  & 11 & 142 & 142 & 142.00 & (0.00) & 0.04 & (0.00) \\
 & 20 & 220 & 220 & 220.05 & (0.22) & 57.07 & (43.23) 			    &  & 12 & 130 & 130 & 130.00 & (0.00) & 0.07 & (0.01) \\
 & 21 & 208 & 208 & 208.00 & (0.00) & 0.12 & (0.00) 				    &  & 13 & 120 & 120 & 120.00 & (0.00) & 0.08 & (0.01) \\
 & 22 & 200 & 200 & 200.00 & (0.00) & 0.13 & (0.06) 				    &  & 14 & 111 & 111 & 111.00 & (0.00) & 0.29 & (0.24) \\
 & 23 & 189 & 189 & 189.00 & (0.00) & 0.12 & (0.02) 				    &  & 15 & 104 & 104 & 104.00 & (0.00) & 0.27 & (0.13) \\
 & 24 & 179 & 179 & 179.00 & (0.00) & 0.48 & (0.41) 				    &  & 16 & 98 & 98 & 98.00 & (0.00) & 0.08 & (0.01) \\
 & 25 & 172 & 172 & 172.00 & (0.00) & 0.43 & (0.18) 				    &  & 17 & 92 & 92 & 92.00 & (0.00) & 0.26 & (0.14) \\
 & 26 & 171 & 171 & 171.00 & (0.00) & 0.27 & (0.03)  			    &  & 18 & 87 & 87 & 87.00 & (0.00) & 0.27 & (0.14) \\ \cline{1-8}
Sawyer (30) & 7 & 47 & 47 & 47.00 & (0.00) & 0.14 & (0.11) 			    &  & 19 & 84 & 84 & 84.00 & (0.00) & 0.10 & (0.06) \\
 & 8 & 41 & 41 & 41.00 & (0.00) & 0.06 & (0.03) 				    &  & 20 & 79 & 79 & 79.00 & (0.00) & 0.13 & (0.03) \\
 & 9 & 37 & 37 & 37.00 & (0.00) & 0.03 & (0.01) 				    &  & 21 & 76 & 76 & 76.00 & (0.00) & 0.08 & (0.04) \\
 & 10 & 34 & 34 & 34.00 & (0.00) & 0.02 & (0.00) 				    &  & 22 & 73 & 73 & 73.00 & (0.00) & 0.53 & (0.36) \\
 & 11 & 31 & 31 & 31.00 & (0.00) & 0.82 & (0.85) 				    &  & 23 & 69 & 69 & 69.00 & (0.00) & 18.10 & (21.25) \\
 & 12 & 28 & 28 & 28.00 & (0.00) & 0.03 & (0.01) 				    &  & 24 & 66 & 66 & 66.90 & (0.31) & 2.88 & (9.10) \\
 & 13 & 26 & 26 & 26.00 & (0.00) & 0.02 & (0.00) 				    &  & 25 & 64 & $^{\ast}$65 & 65.00 & (0.00) & 12.30 & (11.87) \\
 & 14 & 25 & 25 & 25.00 & (0.00) & 0.02 & (0.00)  			    &  & 26 & 64 & 64 & 64.00 & (0.00) & 0.05 & (0.01) \\ \cline{1-8}
Scholl (297) & 25 & 2787 & 2787 & 2787.00 & (0.00) & 37.11 & (6.11) 		    &  & 27 & 60 & 60 & 60.00 & (0.00) & 1.61 & (0.73) \\
 & 26 & 2680 & 2680 & 2680.00 & (0.00) & 32.56 & (4.71) 			    &  & 28 & 59 & 59 & 59.00 & (0.00) & 0.16 & (0.06) \\
 & 27 & 2580 & 2580 & 2580.00 & (0.00) & 58.06 & (8.65) 			    &  & 29 & 56 & 56 & 56.00 & (0.00) & 1.43 & (1.15) \\ \cline{9-16}
 & 28 & 2488 & 2488 & 2488.00 & (0.00) & 53.78 & (12.43) 			    & Wee-Mag (75) & 3 & 500 & 500 & 500.00 & (0.00) & 0.01 & (0.00) \\
 & 29 & 2402 & 2402 & 2402.00 & (0.00) & 45.57 & (9.93) 			    &  & 4 & 375 & 375 & 375.00 & (0.00) & 0.03 & (0.00) \\
 & 30 & 2322 & 2322 & 2322.30 & (0.47) & 96.86 & (53.36) 			    &  & 5 & 300 & 300 & 300.00 & (0.00) & 0.02 & (0.00) \\
 & 31 & 2247 & $^{\ast}$2248 & 2248.00 & (0.00) & 51.23 & (4.32) 		    &  & 6 & 250 & 250 & 250.00 & (0.00) & 0.03 & (0.00) \\
 & 32 & 2177 & 2177 & 2177.50 & (0.51) & 91.50 & (63.17) 			    &  & 7 & 215 & 215 & 215.00 & (0.00) & 0.01 & (0.00) \\
 & 33 & 2111 & 2111 & 2111.80 & (0.41) & 57.19 & (39.95) 			    &  & 8 & 188 & 188 & 188.00 & (0.00) & 0.04 & (0.01) \\
 & 34 & 2049 & 2049 & 2049.20 & (0.41) & 117.47 & (53.03) 			    &  & 9 & 167 & 167 & 167.00 & (0.00) & 0.01 & (0.00) \\
 & 35 & 1991 & 1991 & 1991.00 & (0.00) & 95.54 & (17.19) 			    &  & 10 & 150 & 150 & 150.00 & (0.00) & 0.08 & (0.02) \\
 & 36 & 1935 & $^{\ast}$1936 & 1936.00 & (0.00) & 61.22 & (8.78) 		    &  & 11 & 137 & 137 & 137.00 & (0.00) & 0.06 & (0.01) \\
 & 37 & 1883 & 1883 & 1883.05 & (0.22) & 140.57 & (34.46) 			    &  & 12 & 125 & 125 & 125.00 & (0.00) & 0.10 & (0.02) \\
 & 38 & 1834 & 1834 & 1834.00 & (0.00) & 125.92 & (34.53) 			    &  & 13 & 116 & 116 & 116.00 & (0.00) & 0.02 & (0.00) \\
 & 39 & 1787 & 1787 & 1787.35 & (0.49) & 192.12 & (55.85) 			    &  & 14 & 108 & 108 & 108.00 & (0.00) & 0.54 & (0.60) \\
 & 40 & 1742 & 1742 & 1742.00 & (0.00) & 153.02 & (29.66) 			    &  & 15 & 100 & 100 & 100.00 & (0.00) & 0.24 & (0.19) \\
 & 41 & 1700 & 1700 & 1700.00 & (0.00) & 81.51 & (18.06) 			    &  & 16 & 94 & 94 & 94.00 & (0.00) & 0.05 & (0.01) \\
 & 42 & 1659 & $^{\ast}$1660 & 1660.00 & (0.00) & 146.58 & (32.11) 		    &  & 17 & 89 & 89 & 89.00 & (0.00) & 0.42 & (0.28) \\
 & 43 & 1621 & 1621 & 1621.00 & (0.00) & 149.08 & (21.75) 			    &  & 18 & $\overline{87}$ & 87 & 87.00 & (0.00) & 0.07 & (0.01) \\
 & 44 & 1584 & 1584 & 1584.15 & (0.37) & 127.52 & (54.06) 			    &  & 19 & $\overline{85}$ & 85 & 85.00 & (0.00) & 0.08 & (0.02) \\
 & 45 & 1549 & 1549 & 1549.00 & (0.00) & 150.65 & (32.59) 			    &  & 20 & 77 & 77 & 77.00 & (0.00) & 0.13 & (0.04) \\
 & 46 & 1515 & $^{\ast}$1516 & 1516.00 & (0.00) & 121.28 & (20.06) 		    &  & 21 & 72 & 72 & 72.00 & (0.00) & 0.08 & (0.01) \\
 & 47 & 1483 & $^{\ast}$1484 & 1484.00 & (0.00) & 165.44 & (28.64) 		    &  & 22 & 69 & 69 & 69.00 & (0.00) & 0.06 & (0.01) \\
 & 48 & 1452 & 1452 & 1452.70 & (0.47) & 151.98 & (60.76) 			    &  & 23 & $\overline{67}$ & 67 & 67.00 & (0.00) & 4.34 & (3.58) \\
 & 49 & $\overline{1427}$ & \colorbox{lightgray}{1424} & 1424.00 & (0.00) & 113.46 & (22.47)   &  & 24 & 66 & 66 & 66.00 & (0.00) & 0.25 & (0.14) \\
 & 50 & 1394 & 1394 & 1394.85 & (0.37) & 143.63 & (53.48) 			    &  & 25 & 65 & 65 & 65.00 & (0.00) & 11.36 & (26.58) \\
 & 51 & 1386 & 1386 & 1386.00 & (0.00) & 2.23 & (0.09) 			    &  & 26 & 65 & 65 & 65.00 & (0.00) & 0.05 & (0.01) \\
 & 52 & 1386 & 1386 & 1386.00 & (0.00) & 0.17 & (0.01)  		    &  & 27 & $\overline{65}$ & 65 & 65.00 & (0.00) & 0.02 & (0.00) \\ \cline{1-8}
Tonge (70) & 3 & 1170 & 1170 & 1170.00 & (0.00) & 0.01 & (0.00) 			    &  & 28 & $\overline{64}$ & 64 & 64.00 & (0.00) & 0.02 & (0.00) \\
 & 4 & 878 & 878 & 878.00 & (0.00) & 0.02 & (0.00) 				    &  & 29 & 63 & 63 & 63.00 & (0.00) & 0.02 & (0.00) \\
 & 5 & 702 & 702 & 702.00 & (0.00) & 0.05 & (0.02) 				    &  & 30 & 56 & 56 & 56.00 & (0.00) & 0.07 & (0.00) \\ \hline
\end{tabular}}
\end{center}
\end{table}

\end{appendix}

\end{document}